\PassOptionsToPackage{pdfpagelabels=false}{hyperref}
\documentclass[fleqn,usenatbib,useAMS]{mnras}
\usepackage{graphicx}
\usepackage{amsmath} 
\usepackage{amssymb}
\usepackage[]{url}
\usepackage{hyperref}
\hypersetup{final}
\usepackage{multirow}
\usepackage{tabularx}
\usepackage{arydshln}
\usepackage{widetext}

\def \prd {PRD}

\def \apj {ApJ}
\def \apjs {ApJS}
\def \apjl {ApJL}
\def \mnras {MNRAS}
\def \aap {A\&A}
\def \aj {AJ}
\def \araa {ARAA}

\def \nat {Nature}

\def \nar {New. A. Rev.}

\def\deg{$^\circ$}

\def\t9{T$_{90}$}

\def\deg{$^{\circ}$}
\newcommand{\msol}{$M_{\odot}$}

\newcommand{\rate}{$\mathrm{Gpc}^{-3}\, \mathrm{yr}^{-1}$}

\newcommand{\pflux}{$ \mathrm{ph}\, \mathrm{sec}^{-1}\, \mathrm{cm}^{-2}$}
\newcommand{\lum}{$ \mathrm{erg}\, \mathrm{sec}^{-1}$}

\newcommand\R{\rule{0pt}{3.5ex}} 

\title[]{Joint gravitational wave - gamma-ray burst detection rates in the aftermath of GW170817}

\author[]{E. J. Howell$^{1}$\thanks{E-mail:eric.howell@uwa.edu.au}, K. Ackley$^{2}$, A. Rowlinson$^{3,4}$  and D. Coward$^{1}$\\
$^1$\,OzGrav-UWA, School of Physics and Astrophysics, University of Western Australia, Crawley WA 6009, Australia\\
$^2$\,OzGrav-Monash Centre for Astrophysics, School of Physics and Astronomy,
Monash University, Clayton, VIC 3800, Australia\\
$^3$\,Anton Pannekoek Institute, University of Amsterdam, Science Park 904, 1098 XH Amsterdam, The Netherlands\\
$^4$\,ASTRON, The Netherlands Institute for Radio Astronomy, Postbus 2, 7990 AA, Dwingeloo, The Netherlands
}

\begin{document}
\linespread{1.0}
\maketitle
\begin{abstract}
The observational follow-up campaign of the gravitational wave (GW) multi-messenger event GW170817/GRB170817A has shown that the prompt $\gamma$-rays are consistent with a relativistic structured jet observed from a wide viewing angle $\gtrsim 20$\deg. We perform Bayesian inference using the data from early and late EM observations to determine the jet profile of GRB170817A assuming a structured jet model. We use the geometric dependence on the burst luminosity to produce a short duration gamma-ray burst (sGRB) efficiency function with redshift, which folded in with binary neutron star detection rate, allows us to estimate the future joint GW/sGRB detection rates for LIGO and Virgo detectors.
We show that, if the jet structured profile of GRB170817A is a relatively common feature of sGRBs, then there is a realistic probability of another off-axis coincident detection during the third aLIGO/Virgo observing run (O3). We also find that up to 4 yr$^{-1}$ joint events may be observed during the advanced LIGO run at design sensitivity and up to 10 yr$^{-1}$ by the upgraded advanced LIGO configuration A+. We show that the detection efficiencies for wide-angled sGRB emissions will be limited by GRB satellites as the GW detection range increases through proposed upgrades. Therefore, although the number of coincident detections will increase with GW detector sensitivity, the relative proportion of detected binary neutron stars with $\gamma$-ray counterparts will decrease; 11\% for O3 down to 2\% during A+.
\end{abstract}

\begin{keywords}
gamma-rays: bursts -- gamma-ray: observations -- methods: data analysis -- cosmology: miscellaneous
\end{keywords}


\graphicspath{{/figs/}{./}}

\section{Introduction}
The first gravitational wave observation of a system of coalescing binary neutron stars (BNSs) GW170817, and the coincident detection by Fermi \citep{goldstein_ordinary_2017} and INTEGRAL \citep{savchenko_integral_2017} of a short-duration gamma-ray burst (sGRB) within 1.7\,s, firmly established that these two types of events are associated \citep[][]{LSC_GW_GRB_2017ApJ}. The subsequent localization to within 28\,deg$^2$\,\citep{LSC_BNS_2017PhRvL} prompted a ground breaking electromagnetic follow-up campaign \citep{LSC_MM_2017ApJ,Alexander_2017ApJ,Andreoni_2017PASA,Hallinan_2017Sci,Kasliwal_2017Sci,Margutti_2017ApJ,Troja_2017Natur} providing further proof of the association via the detection of a macronova/kilonova; a thermal afterglow powered through the radioactive decay of heavy elements produced through rapid neutron capture in the material ejected during the violent NS merger \citep[]{Li1998ApJL, Metzger2010MNRAS}.

Although predictions for the association of sGRBs with the merger of NSs have existed for several decades \citep[]{Eichler1989Nature}, the relative close proximity of GRB170817A and the dimness of the prompt gamma-ray emission was unexpected. The occurrence of a BNS merger at 40\,Mpc was not unreasonable considering upper rate predictions of \citet{Abadie2010CQGra,LSC_BNS_2017PhRvL}. However, given that the small sample of sGRBs with well determined opening angles\footnote{We note that there are only 4 sGRBs with observed jet breaks which allow unambiguous opening angle determinations.} are in the range $\theta_{j}\sim$4--8\deg \citep{fong_decade_2015} the detection of a sGRB within such a volume was unexpected. Only one burst with a known redshift has been detected closer \citep[GRB\,980425 at $d_L \sim$ 35\,Mpc;][]{galama_unusual_1998,woosley_gamma-ray_1999} and it was classified as a low-luminosity long-duration GRB; the event rates of this population are estimated to be least an order of magnitude greater \citep{2007A&A...465....1D,2007ApJ...657L..73G,Howell2013MNRAS} than most predictions of sGRBs \citep{coward_swift_2012}.

In terms of the energetics of GRB170817A, given the close proximity and relatively standard gamma-ray flux, the resulting luminosity of $\sim 10^{47}$ \lum\, is 2 orders of magnitude below the generally accepted lower limits on the sGRB luminosity function and several orders of magnitude below the average luminosity for sGRBs \citep[see for example the sample of][]{WandermanPiran2015MNRAS}.\\
\indent
A resolution was provided by careful observations from radio through to X-ray.
Follow-up by the X-ray telescope (XRT) on the Neil Gehrels Swift Observatory (\emph{Swift}) after 0.5\,d and by the Nuclear Spectroscopic Telescope Array
(NuSTAR) after 0.6\,d yielded upper limits \citep{evans_swift_2017} which strongly suggested the $\gamma$-ray emission was not viewed along the jet axis. These findings were supported by further monitoring in the radio \citep{ruan_brightening_2018,resmi_low_2018}, optical \citep{lyman_optical_2018} and X-ray bands \citep{margutti_binary_2018,davanzo_evidence_2018,troja_outflow_2018} which suggested Fermi/INTEGRAL detected the prompt emission from a wide-angle \mbox{($\theta_{j}\sim20$\deg--\,$30$\deg)}, which would be significantly weaker than one viewed along the jet axis.\\
\indent Although the implications of wider angled prompt emissions and delayed afterglows had been discussed by a number of studies \citep[e.g.][]{hotokezaka_mass_2015,nakar_observable_2016,evans_optimization_2016,Lazzati2017ApJ}, previous to GRB170817A, joint sGRB-GW event rates had been based on $\gamma$-ray emissions produced within highly collimated sub-relativistic outﬂows \citep{coward_swift_2012,clark_prospects_2015,regimbau_revisiting_2015}. This \emph{top-hat} model assumption significantly reduced the possibility of joint GW/sGRB detections based on the small sample of observations suggesting $\theta_{j}\lesssim$10\deg.\\
\indent There are currently two leading models to produce a wide angled gamma-ray emission: a \emph{structured jet} in which the luminosity per unit solid angle decreases smoothly with an angular dependence outside a narrow core \citep{Kathirgamaraju_2018MNRAS} and a hot expanding mildly-relativistic \emph{cocoon}, produced by an injection of energy into the post-merger circum-burst ejector by a successful or choked jet \citep{hotokezaka_mass_2015,gottlieb_cocoon_2017}. Studies suggest that a structured jet could be a by product of a successful jet penetrating a cocoon \citep{lazzati_late_2017} and have also been termed \emph{successful structured jets} \citep{Alexander_2018ApJ}; thus the structured jet model is appropriate to both scenarios.\\
\indent Continued late-time monitoring of GRB170817A ($>$ 200 days) from radio to X-ray are veering heavily in support of the successful jet scenario. Recent long baseline interferometric (VLBI) observations of GW170817/GRB170817A showed super-luminal motion, demonstrating that a successful jet core with an opening angle of $<5$\deg was launched and that the early emissions were from a successful structured jet viewed 20\deg from the jet axis \citep{Mooley_2018}. These observations have been supported by \citet{Alexander_2018ApJ}. We are motivated therefore to assume a structured jet model to estimate future joint GW/sGRB event rates.\\

\indent In this paper we will present joint GW/sGRB event rates for future LIGO/Virgo observation runs using the most probable structured jet profile of GRB170817A and assuming that all BNS mergers can in principle produce a sGRB\footnote{Noting that the actual detectability of the sGRB depends on the viewing angle and energetics.}. We note that although black hole - neutron star mergers (BH/NS) are also considered as progenitors of sGRBs \citep{pannarale_prospects_2014,Bhattacharya_2018}, we base this study around the firm BNS/sGRB association resulting from GW170817/GRB170817A.. We will first (in Section 2) infer the most probable jet profile of GRB170817A based on electromagnetic follow-up observations and informed observation based priors. We will then present a theoretical framework for determining joint GW/sGRB detection rates (Section 3) and go on to present detection rate estimates for the current (O3) and design sensitivity Advanced LIGO \citep{aLIGO_2015} and Advanced Virgo \citep{Acernese_2015} configurations and for enhanced aLIGO configurations, A+ \citep{Barsotti_Aplus_2018} and Voyager \citep{Voyager_2018} (Section 4). Finally, we conclude (Section 5) by discussing the implications of our findings.

\section{Inferring the Structured Jet profile}
\subsection{The Structured Jet profile}
A structured jet has an angular dependence on energy and bulk Lorentz factor, and is generally described by an ultra-relativistic core without sharp edges that smoothly transforms to a milder relativistic outflow at greater angles \citep{lipunov_gamma-ray_2001,rossi_afterglow_2002,zhang_gamma-ray_2002,nakar_testing_2004,lamb_electromagnetic_2017}. Typical angular profiles are provided by Gaussian or power-law jet models. Given the uncertainty provided by one firm observation, the majority of late-time EM follow up campaigns have considered the former model; we therefore consider this model to describe the angular dependence on energy. A Gaussian structured jet model has the form \citep{zhang_gamma-ray_2002,resmi_low_2018}:
\begin{equation}
L (\theta_{\mathrm{V}}) = L_{\mathrm{c}} \exp \left( -\frac{\theta_{\mathrm{V}}^{2}}{2\theta^{2}_{\mathrm{c}}} \right)
 \,, \label{eq_structure_gaussian}
\end{equation}
\noindent with $L (\theta)$ the luminosity per unit solid angle, $\theta_{\mathrm{V}}$ the viewing angle and  $L_{\mathrm{c}}$ and $\theta_{\mathrm{c}}$ structure parameters that define the sharpness of the angular profile. To reproduce the structured jet profile for GRB170817A one requires the best estimates of both these two parameters.

\subsection{Structured Jet parameters inferred from late time observations}
Table \ref{table_sj_observations} documents the varied range of estimates of $\theta_{\mathrm{c}}$ and the viewing angle $\theta_{\mathrm{v}}$ obtained through multi-wavelength observations of the late time emissions of GRB170817A up to $\sim$150 days; such modelling also considers the isotropic kinetic energy of the jet $E_{\mathrm{K,ISO}}$, the circum-burst particle density $n$, the microphysical parameters, $\epsilon_{\mathrm{B}}$ and $\epsilon_{\mathrm{e}}$ that describe the fractions of post-shock energy in the magnetic fields and radiating electrons respectively and the electrons energy power law distribution index $p$. Due to degeneracies that exist in these parameters, i.e. $\theta_{\mathrm{v}}$ is found to correlate strongly with $n$ and $\theta_{\mathrm{c}}$ and anti-correlate with $E_{\mathrm{K,ISO}}$ \citep{troja_outflow_2018,lazzati_late_2017} there are large uncertainties on the models; this has been most apparent through the spread in estimates of $\theta_{\mathrm{v}}$.

\begin{figure}
 \centering
 \includegraphics[scale = 0.64,origin=rl]{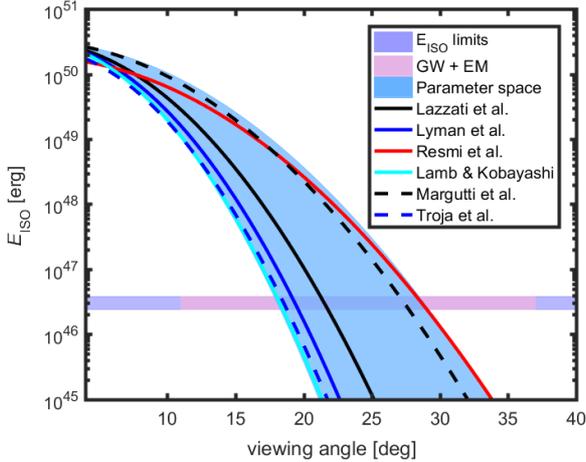}
 \caption{Curves of  E$_{\mathrm{ISO}}$ as a function of viewing angle are shown for the Gaussian jet models provided in Table \ref{table_sj_observations}. Curves assume an efficiency $\eta_{K,\gamma}$ of 0.4 as defined in equation \ref{eq_ke_efficiency}. The horizontal band shows the Fermi-GBM E$_{\mathrm{ISO}}$ limits; within this band we also show the EM constraints on beaming angle from \citep{finstad_measuring_2018}. The shaded diagonal portion indicates the parameter space on structured jets given in Table \ref{table_sj_observations} with viewing angles ranging from 18 \deg to 33 \deg.
 }
  \label{fig:structured_models}
\end{figure}

Figure \ref{fig:structured_models} shows the isotropic equivalent energy $E_{\mathrm{\gamma,ISO}}$ per unit solid angle as a function of viewing angle $\theta_{\mathrm{v}}$ using the structured jet model parameters provided in \mbox{Table \ref{table_sj_observations}}. To compare with the Fermi-Gamma-ray Burst Monitor \citep[Fermi-GBM,][]{FermiGBM2009} prompt observations, all curves assume the isotropic equivalent kinetic energy of the jet $E_{\mathrm{K,ISO}}$ is converted to $E_{\mathrm{\gamma,ISO}}$ with an efficiency:
\begin{equation}
  \eta_{K,\gamma} = \frac{E_{\mathrm{\gamma,ISO}}}{E_{\mathrm{\gamma,ISO}} + E_{\mathrm{K,ISO}} }\,.
  \label{eq_ke_efficiency}
\end{equation}
\noindent For an estimate of $\eta_{K,\gamma}$ we follow \citet{fong_decade_2015} who find a median value of $\eta_{K,\gamma} = 0.4$ assuming $\epsilon_{\mathrm{B}}=0.01$. To produce a set of curves we adopt this value in the equation (\ref{eq_structure_gaussian}) along with the documented value of $\theta_{\mathrm{c}}$. Figure \ref{fig:structured_models} shows the Fermi-GBM observed $E_{\mathrm{\gamma,ISO}}$ as the horizontal band; the point each model crosses this band is the inferred viewing angle of the prompt emission. The plot shows that a range of models predict values of $\theta_{\mathrm{v}}$ within the range 18-33\deg.
\begin{table*}
  \caption{ The model parameters for Gaussian jet models used to fit to the late EM follow-up data of GW170817/GRB170817A. Unless stated otherwise we assume the model given in equation (\ref{eq_structure_gaussian}). KEY: \textbf{Radio:} GMRT-Giant Metrewave Radio Telescope; VLA- Karl G. Jansky Very Large Array; ATCA- Australian Telescope Compact Array; uGMRT - upgraded Giant Metrewave Radio Telescope; \textbf{Optical/NIR:} HST - Hubble Space telescope; SSS - Swope Supernova Survey \textbf{X-ray:} CXO-Chandra X-ray Observatory; NuSTAR - Nuclear Spectroscopic Telescope ARray; XMMN-XMM–Newton  }
\begin{center}
  \begin{tabular}{lllllll}
\hline
Study  &Core $\theta_{\mathrm{c}}$ & $E_{\mathrm{K,iso}}$ & viewing angle $\theta_{\mathrm{v}}$  & Instrument/Data &  observations    \\
       &    [deg]                 & $\times 10^{52}$ erg                 & [deg]          &                 & (days)
\\
\hline
\hline
\citet{lazzati_late_2017}  & 5    & 1 & 22  & ATCA;VLA; HST;SSS; uGMRT; CXT & 16 - 145
\\
\citet{lyman_optical_2018} & 4.5  & 1 &  19    & HST                           & 5-11,110
\\
\citet[][]{resmi_low_2018} $\dagger$ & 6.9 & 0.58 & 29    & GMRT &  7-152
\\
\citet{lamb_grb_2017} $\ddagger$     & 6   & 1    & 18    &  Fermi-GBM; INTEGRAL    & Prompt emission
\\
\citet{margutti_binary_2018} $\ddagger$ & 9 & 1 & 27  & GMRT  &  7-152
\\
\citet{troja_outflow_2018}   & 4.4   & 3.3  & 23  & CXT; NuSTAR; ATCA & 9 - 110
\\
\hline
\end{tabular}

$\dagger$ \citet{gill_afterglow_2018} also derived similar results to this study. $\ddagger$ based on a modified Gaussian jet model with $\epsilon^{\theta^{2}/\theta_{\mathrm{c}}^{2}}$.
  \label{table_sj_observations}
  \end{center}
\end{table*}

Tighter constraints on $\theta_{\mathrm{v}}$ have now been provided through VLBI observations of GW170817/GRB170817A that demonstrated a successful narrow relativistic jet was launched and an associated viewing angle of $20\pm5$\deg \citep{Mooley_2018}. These findings have been supported by radio, optical and X-ray data of \citet{Alexander_2018ApJ}. In the remainder of this section we will adopt the observations of \citep{Mooley_2018} in our priors to estimate the most likely structured jet profile for GRB170817A. We will use this profile to calculate a sGRB detection efficiency function which will be used to determine joint sGRB/GW rate estimates; we note that we will verify this function through agreement with the number of sGRBs observed each year by Fermi-GBM.

\subsection{Inferring the structured jets profile of GRB170817A}
\label{sec:jet_profile_inference}

To infer the parameters of the structured jet profile of GRB170817A given the observations $\mathrm{L}_{\mathrm{obs}}$ we use a standard Bayesian framework in which the joint posterior distribution is given as:
\begin{equation}\label{eq_bayes}
 \mathcal{P}(\hat{\theta}\, |\,\mathrm{L}_{\mathrm{obs}}) =  \frac{ \mathcal{L}(\mathrm{L}_{\mathrm{obs}}\, |\, \hat{\theta}\,) \, \mathcal{P}(\hat{\theta})}{\mathcal{Z}}\,,
\end{equation}
\noindent where $\mathcal{L}(\mathrm{L}_{\mathrm{obs}}\, |\, \hat{\theta}\,)$ is the likelihood distribution of the observed luminosity $\mathrm{L}_{\mathrm{obs}}$ given the parameters $\hat{\theta}$, $\mathcal{P}(\hat{\theta})$ are the prior distributions on $\hat{\theta}$ and $\mathcal{Z}$ is a factor termed the \emph{Bayesian evidence}; as this factor enters equation \ref{eq_bayes} as a normalization factor independent of the model parameters, if one is only interested in the posterior distribution rather than model selection, this term can be ingnored.

The likelihood function to infer the Gaussian jet profile is based on the observed luminosity and assumes the parametrisation given by equation \ref{eq_structure_gaussian} and a log-normal likelihood function based on the observed luminosity, $\mathrm{L}_{\mathrm{obs}}$, of GRB170817A \citep{lognormal_1996ApJ}:
\begin{equation}\label{loglike}
  \mathcal{L}(\mathrm{L}_{\mathrm{obs}}\, |\, \hat{\theta}\,) = \frac{1}{\sqrt{2 \pi \sigma^{2}} }
  \,\mathrm{exp}\left[  - \frac{[\,\mathrm{log} \,\mathrm{L}_{\mathrm{obs}} - \mathrm{L}(\hat{\theta} )\,]^{2}}{2\sigma^{2}} \right]\,.
\end{equation}
\noindent Here $\tilde{\mathrm{L}}$ is the observed isotropic equivalent luminosity of GRB170817A and $L(\theta)$ is the luminosity given parameters $\theta = [\theta_{\mathrm{c}},L_{\mathrm{c}},\theta_{\mathrm{v}} $ ]; $\sigma$ is the standard deviation on $\mathrm{L}_{\mathrm{obs}}$.  Using the parameters for the brightest part of burst modeled by a Comptonized spectrum with power law index $\alpha = -0.85$ and E$_{\mathrm{peak}}$=229\,keV and the 64ms peak flux of $(7.3 \pm 2.5 ) \times 10^{-7}$ erg s$^{-1}$ cm$^{-2}$ \citep{goldstein_ordinary_2017}, we find an observed luminosity of $(1.7 \pm 0.6 ) \times 10^{47}$ erg s$^{-1}$ (using equation \ref{eq_flux}) and an associated maximum detection distance of 67 Mpc.

Given this likelihood function we use the \verb"emcee" implementation of the affine-invariant ensemble sampler \citep{ForemanMackey2013} to
execute a MCMC analysis to determine the posteriors of the
model parameters. We use a uniform flat prior distribution for $\theta_{\mathrm{c}}$ in the range 1-9\deg and a Gaussian distribution around $\theta_{\mathrm{v}} = 20\pm5 $ based on the observations of \citep{Mooley_2018}.

For the prior on $L_{\mathrm{c}}$, we use the fact that the maximum detection distance of GRB170817A is relatively local in comparison with known cosmological GRB redshifts. Therefore, it is a reasonable assumption that the majority of sGRBs have been observed at small viewing angles close to the jet core. A reasonable prior is therefore a lognormal distribution with a mean observed sGRB isotropic equivalent luminosity \mbox{$<L_{\mathrm{ISO}}> \approx 2 \times 10^{52}$\lum} which we take from \citet{WandermanPiran2015MNRAS}. We note that the structured jet is generally parameterised using the luminosity per unit solid angle $\mathrm{d}L_{\mathrm{c}}/\Omega$; as the reported luminosity is the isotropic equivalent, we assume the conversion $4\pi\mathrm{d}L_{\mathrm{c}}/\Omega$ for comparison with the observed value $\mathrm{L}_{\mathrm{obs}}$.

We note here that this assumption assumes that GRB170817A would not have been exceptionally different to the previously observed sGRB population if viewed directly along the jet axis. It is known that GRB170817A demonstrated a two-component structure; a short hard spike followed by a longer soft tail possibly thermal in origin, suggesting it was relatively unique burst \citep{gottlieb_cocoon_2017}. A recent study by \citet{Burns_2018ApJ} has focussed on another sGRB, GRB150101B observed at $z = 0.134$ (the third closest GBM observed sGRB) with a similar two component structure to that of GRB170817A. This study proposes that such a two-component structure could be an intrinsic feature of sGRBs only observable at low-$z$ and that 150101B could represent a more on-axis but more distant version of GRB170817A.

\begin{figure}
 \centering
 \includegraphics[scale = 0.16,origin=rl]{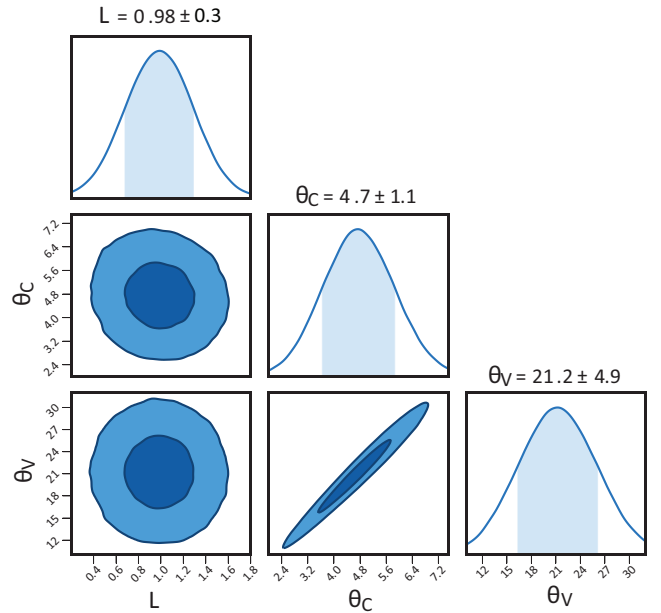}
 \caption{ Posterior distributions for the parameters of a Gaussian structured jet profile of GRB170817A obtained from a MCMC analysis; the shaded regions represent the $1\sigma$ credible intervals. Luminosity is presented as an isotropic equivalent value ($\mathrm{d}L_{\mathrm{c}}\Omega$). Contours in the
two-dimensional posteriors represent the $1\sigma$ and $2\sigma$ confidence intervals.}
  \label{fig:structured_corner}
\end{figure}

Figure \ref{fig:structured_corner} shows the posterior distributions of our MCMC analysis on the structured jet profile parameters of GRB170817A. We find a mean acceptance value of around 35\% on our MCMC chains; we use 30,000 samples and find an integrated auto-correlation time of around 32 yielding around 19,000 independent samples\footnote{Based on the iterative procedure described in
\url{http://www.stat.unc.edu/faculty/cji/Sokal.pdf}}. We find a structured jet profile defined by parameters $ L_{\mathrm{c}} = 1.0\pm 0.3 \times 10^{52}/\Omega$ erg s$^{-1}$ sr$^{-1}$, $\theta_{\mathrm{c}} = 4.7 \pm 1.1 $\deg ($0.08 \pm 0.02$ rad) and a viewing angle of $\theta_{\mathrm{v}} = 21.2 \pm 4.9$\deg ($0.37 \pm 0.09$ rad). The data clearly shows a degeneracy between $\theta_{\mathrm{c}}$ and $\theta_{\mathrm{v}}$; more compact cores produce emissions only observable at smaller viewing angles.

Our estimates of $\theta_{\mathrm{c}}$ are in good agreement with \citep{Mooley_2018} who suggest a strong constraint of $\theta_{\mathrm{c}}<5$\deg.
Additionally, we note that \citet{lamb_grb_2017} also provided estimates of a Gaussian Structured jet using the prompt data from Fermi-GBM and INTEGRAL -- by fine tuning their initial parameter values (provided in Table \ref{table_sj_observations}) they found best estimates for a Gaussian Structured jet with $\theta_{\mathrm{c}} = 4.5$\deg viewed at 20\deg; these values are in close agreement with the values found in this study.

\begin{figure}
 \centering
 \includegraphics[scale = 0.6,origin=rl]{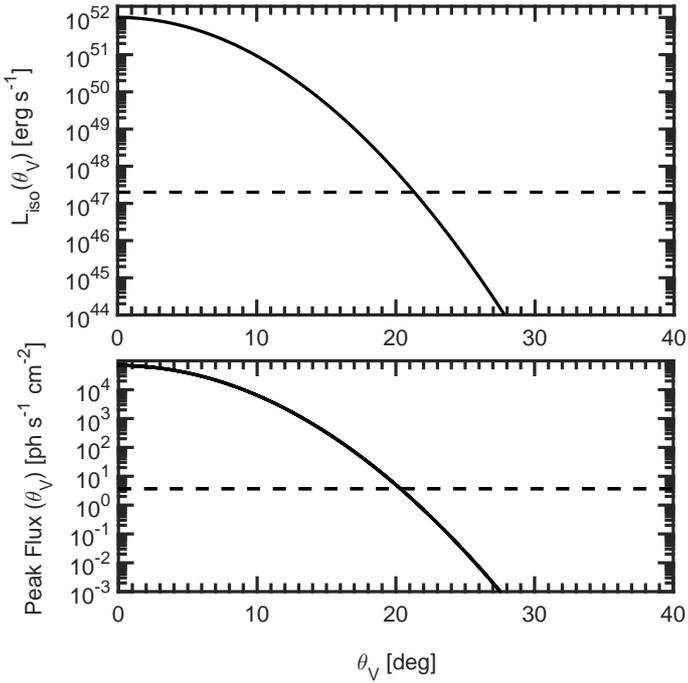}
 \caption{The structured jet profile of GRB170817A shown as an observed isotropic equivalent luminosity (top panel) and observed peak flux (lower panel) as a function of viewing angle. The dashed line indicates the observed quantities.}
  \label{fig:structured_profile}
\end{figure}

Figure \ref{fig:structured_profile} plots the structured jet profile of GRB170817A shown as observed isotropic equivalent luminosity ($ 2 \times 10^{47}$ \lum ) and peak flux (3.7 \pflux\footnote{We note that conversion from isotropic equivalent luminosity to peak photon flux would be a factor of 1.7 greater including the correction factor $b$ of equation \ref{eq_bol_corr}; in this plot we omit this correction term to agree with the value given by \citep{LSC_GW_GRB_2017ApJ} using the convention of \citet{Bloom2001AJ} who consider only a $k$-correction term.}) as a function of viewing angle. The plot shows that the burst would have been exceptionally bright in peak-flux had it been viewed on axis by virtue of its close distance.

\subsection{The sGRB detection efficiency function for a structured jet profile}
\label{sec:Fermi_efficiency_function}
Given a model for the structured jet profile of GRB170817A we now produce a detection efficiency function for sGRBs. This function allows us to determine a joint GW/sGRB rate from a BNS detection rate model that we will later derive in section \ref{sec:bnsrates}. In this study we will limit our discussion to the Fermi-GBM \citep{FermiGBM2009} instrument but note the framework could be applied to any other GRB detection instrument given an accurate measure of the flux limit.

In general, by determining the fraction of the sGRB luminosity function (LF) accessible at each increase in redshift \citep{2007NewAR..51..539C} one arrives at a scaling relation that is the detection efficiency function $\Sigma(L)$. This function can be applied to the intrinsic source rate evolution model to yield a detection rate with redshift. However, if assuming a structured jet profile one must also fold in the geometric dependence; sources at higher-$z$ will only be detected if the viewing angle is closer to the core and via versa for low-$z$ sources.

We first require a LF, $\Phi(L)$, and we assume a standard  broken power law of the form:
\begin{equation}
\Phi(L) \propto
\biggl \lbrace{
\begin{array}{ll}
\hspace{1.0mm}\left(L/L_{*}\right) ^{\alpha} \hspace{3.0mm} L < L^{*}\\
\hspace{1.0mm}\left(L/L_{*}\right) ^{\beta}  \hspace{3.0mm} L \geq L^{*}
\end{array} }
\end{equation}

\noindent where $L$ is the isotropic rest frame luminosity in the 1-10000\,keV energy range and $L_{*}$ is a characteristic luminosity that separates the low and high end of the luminosity function and $\alpha$ and $\beta$ are the characteristic slopes describing these regimes, respectively. We follow \citep{WandermanPiran2015MNRAS} and use the values $\alpha = -1.95$, $\beta = -3$ and $L^{*} = 2\times 10^{52}$\lum. We note that the values for the power law indexes are increased by a factor of unity over the values presented in that study to convert from a logarithmic to a linear distribution of luminosities \citep{LSC_GW_GRB_2017ApJ}. We assume a standard low end cutoff in luminosity of $L_{\mathrm{min}} = 10^{49}$\,\lum.

We calculate the efficiency function by integrating over the detectable fraction of the sGRB LF,
\begin{equation}
\Sigma(z) = \int_{0}^{\theta_{\mathrm{v}}(L_{\mathrm{min}})} \sin\theta \mathrm{d}\theta \int_{L_{\mathrm{min}}(\mathrm{P_{T}},z) }^{L_{\mathrm{max}} }\!\! \acute{\Phi}(L) \mathrm{d}L,
\label{eq_grb_eff}
\end{equation}
\noindent where $\acute{\Phi}(L)$ is the normalized\footnote{We note that the GRB luminosity function for GRB sources is often used with a normalization constant to ensure that it integrates to unity over the range of
source luminosities. This means that $\Phi(L)$ has units of inverse luminosity \citep{pm01}} sGRB LF and $L_{\mathrm{max}}$ approximates the limit of accessibility by a detector with some flux threshold $P_{\mathrm{T}}$ to a source at redshift $z$. For equation (\ref{eq_grb_eff}) using Fermi-GBM, we take the limiting peak flux $P_{\mathrm{T}}$ = 1.05\,\pflux in the 50-300 keV band; up to the time of this publication, over 95\% of the bursts are detected in the 64 ms timescale by the Fermi-GBM burst catalogue\footnote{\url{https://heasarc.gsfc.nasa.gov/W3Browse/fermi/fermigbrst.html}} are within this value.
We note that other estimations of the flux limit have been presented in the literature; in a similar application \citep{LSC_GW_GRB_2017ApJ} use a value of 2 \pflux also in the 50-300 keV band -- this slightly higher value was adopted to account for the sky-dependent sensitivity of GBM which we account for later in equation \ref{eq_Rz_GRB} (as $\Lambda_{\mathrm{GRB}}$; the time-averaged observable sky fraction). Our limit is also lower than the value of 2.37 \pflux used by \citet{WandermanPiran2015MNRAS} in the same 50--350 keV energy band; we find that around 80\% of the Fermi-GBM burst are below this value.

To determine $L_{\mathrm{min}}(\mathrm{P_{T}},z) $ in equation(\ref{eq_grb_eff}) we require the observed peak photon flux $P$ [ph cm$^{-2}$ s$^{-1}$] in a detectors sensitive energy window $E_{\mathrm{min}}\hspace{-1.0mm}<\hspace{-1.0mm}E\hspace{-1.0mm}<\hspace{-1.0mm}E_{\mathrm{max}}$ from a source at redshift $z$ to be $P = P_T$. This is given by the standard flux-luminosity relation with two corrections: an energy normalisation and a k-correction \citep{WandermanPiran2015MNRAS}. Firstly, the observed photon flux is scaled by
\begin{equation}\label{eq_bolcorr}
b = \int^{E2}_{E1}E\,S(E)dE / \int^{10000}_{1}E\,S(E)dE\,,
\end{equation}\label{eq_bol_corr}
\noindent to account for the missing fraction of the gamma-ray energy seen in the detector band, where $S(E)$ is the observed GRB photon spectrum in units of ph s$^{-1}$\,keV$^{-1}$\,cm$^{-2}$.

Secondly, we apply a cosmological $k$-correction of the form
\begin{equation}\label{eq_k_corr}
k(z) =  \int^{E2}_{E1}S(E)dE / \int^{E2(1 + z)}_{E1(1 + z)}S(E)dE
\end{equation}\label{eq_kcorr}
\noindent such that the standard definition for the flux from a source at a luminosity distance, $d_{\mathrm{L}}(z)$, becomes
\vspace{-0.5mm}
\begin{equation}\label{eq_flux}
    P =  (1 + z) \frac{L}{4 \pi d_{\mathrm{L}}(z)^{2} } \frac{b}{k(z)} \,.
\end{equation}
\noindent We note that the $(1 + z)$ factor is included as the standard definition of $d_{\mathrm{L}}(z)$ is valid for an energy flux, but we convert to photon counts as defined for $P$ \citep{meszaros_cosmological_2011}.

For sGRBs we model the function $S(E)$ by the Band function \citep{Band03} which is a phenomenological fit to the observed spectra of 	GRB prompt emissions and is a function of spectral indices ($\alpha'$,$\beta'$) and break energy, $E_b$, where the two power laws combine.
We take the parameters of this function from \citet{WandermanPiran2015MNRAS} who for Fermi-GBM sGRB spectra found low and high energy spectral indices of $\alpha'$=-2.25 and $\beta'$=-0.5, respectively, and a peak energy $E_{\mathrm{peak}}$ = 800 keV in the source frame.

\begin{figure}
 \centering
 \includegraphics[scale = 0.66,origin=rl]{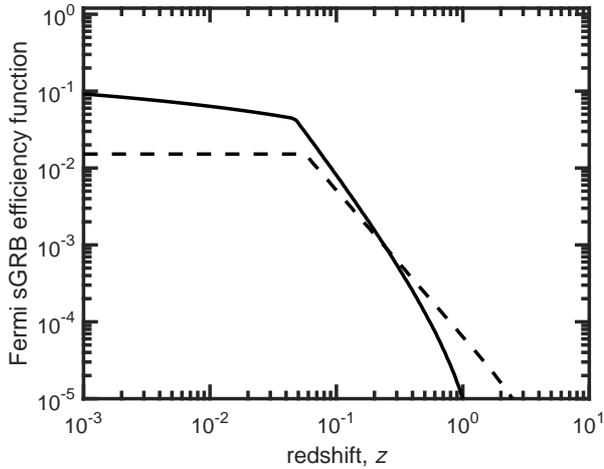}
 \caption{The Fermi-GBM detection efficiency function for a structured jet with a jet profile similar to GRB170817A is shown by the solid curve. For comparison the efficiency function assuming a top hat jet is shown (dashed lines).}
  \label{fig:sgrb_efficiency}
\end{figure}

Figure \ref{fig:sgrb_efficiency} shows the Fermi-GBM detection efficiency function. We note that in this model predicts more detections up to $z\sim0.08$ in comparison to an efficiency function modeled using a top-hat approximation based on a half opening angle of 10\deg (shown with the dashed lines and calculated using the relation shown in Appendix B). This suggests that in the regime $z\sim 0.08$ ($\sim$370 Mpc)wider-angled emissions ($>$10\deg) can be detected. Above $z\sim 0.08$ a top-hat modelled efficiency function would dominate as the wider-angled emissions are no longer above the detection threshold for Fermi-GBM.

\begin{figure}
 \centering
 \includegraphics[scale = 0.66,origin=rl]{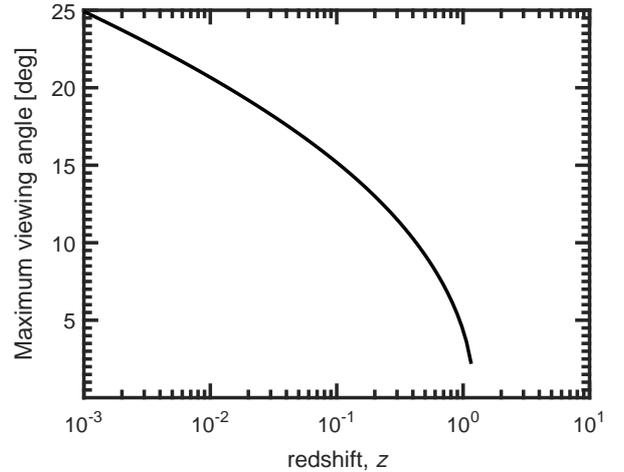}
 \caption{The maximum observable viewing angle as a function of redshift for a sGRB with a structured jet profile like GRB170817A. The plot shows the interplay between detection and opening angle with increasing redshift; at large redshifts sGRBs must be viewed close to the core of a structured jet to enable detection.}
  \label{fig:fermi_max_angle_z}
\end{figure}

Figure \ref{fig:fermi_max_angle_z} illustrates the relation between viewing angle and distance by showing the maximum observable viewing angle as a function of redshift for a sGRB with a structured jet profile like GRB170817A. We see that emissions from viewing angles greater than 10\deg can be detected within $z\sim 0.4$; detections at greater distances will require viewing angles closer to the core. The interplay between viewing angle and detection range is however more complex; at high-$z$ the rate of sources increases whilst at low-$z$ the number of events that can be observed at wider viewing angles are rarer. This latter relationship will become apparent when we calculate detection rates in sections \ref{sec:bns_event_rates} and \ref{sec:Fermi_det_rate}.

\section{Event rates }
\label{sec:bnsrates}
In the following section, we present a framework to model the cosmic BNS source rate evolution and the detection rate of BNS mergers. We convert the intrinsic BNS rate to detection rates by folding in GW detection efficiency functions which will provide a measure of the fraction of sources detected by the instrument with increasing source redshift.

\subsection{The all-sky event rate equation of BNS coalescence}
\label{sect:event_rate}
To estimate the rate of coalescing BNSs as a function of redshift we first assume that the formation rate $R_{\mathrm{BNS}}(z)$ tracks the star formation history of the Universe ($R_{\mathrm{F}}(z)$, in units of $\mathrm{M}_{\odot}\,\mathrm{yr}^{-1}\, \mathrm{Mpc}^{-3}$), with a typical delay time $t_{\mathrm{d}}$ marking the time from binary formation until final merger. For $R_{\mathrm{F}}(z)$ we adopt the extinction-corrected cosmic star formation rate model of  \citet{MadauDickinson2014ARA&A} 
The delay time $t_{\rm d}$ between the formation of the binary system $t_{\rm f}$ and the age of the Universe at the time of merger $t(z)$ is given as
\begin{equation}
t_{d}= \int_{z}^{z_{\mathrm{f}}} \frac{\mathrm{d}z'} {(1+z')H'(z')},
\end{equation}
where $z_{\rm f}$ and $z$ represent the redshifts at which BNS systems form and merge, respectively, and $H'(z) = H_0\sqrt{\Omega_\Lambda + \Omega_\mathrm{m}(1+z)^3}$. For the following calculations we assume a `flat-$\Lambda$' cosmology with the cosmological parameters $\Omega_{\mathrm
m}=0.31$, $\Omega_{\mathrm \Lambda}=0.69$ and
\mbox{$H_{0}=67.8$ km s$^{-1}$ Mpc$^{-1}$} \citep{Planck2015}.

\begin{figure}
 \centering
 \includegraphics[bbllx = 70pt,bblly =5pt, bburx = 350pt, bbury =280pt,scale =0.65,origin=rl]{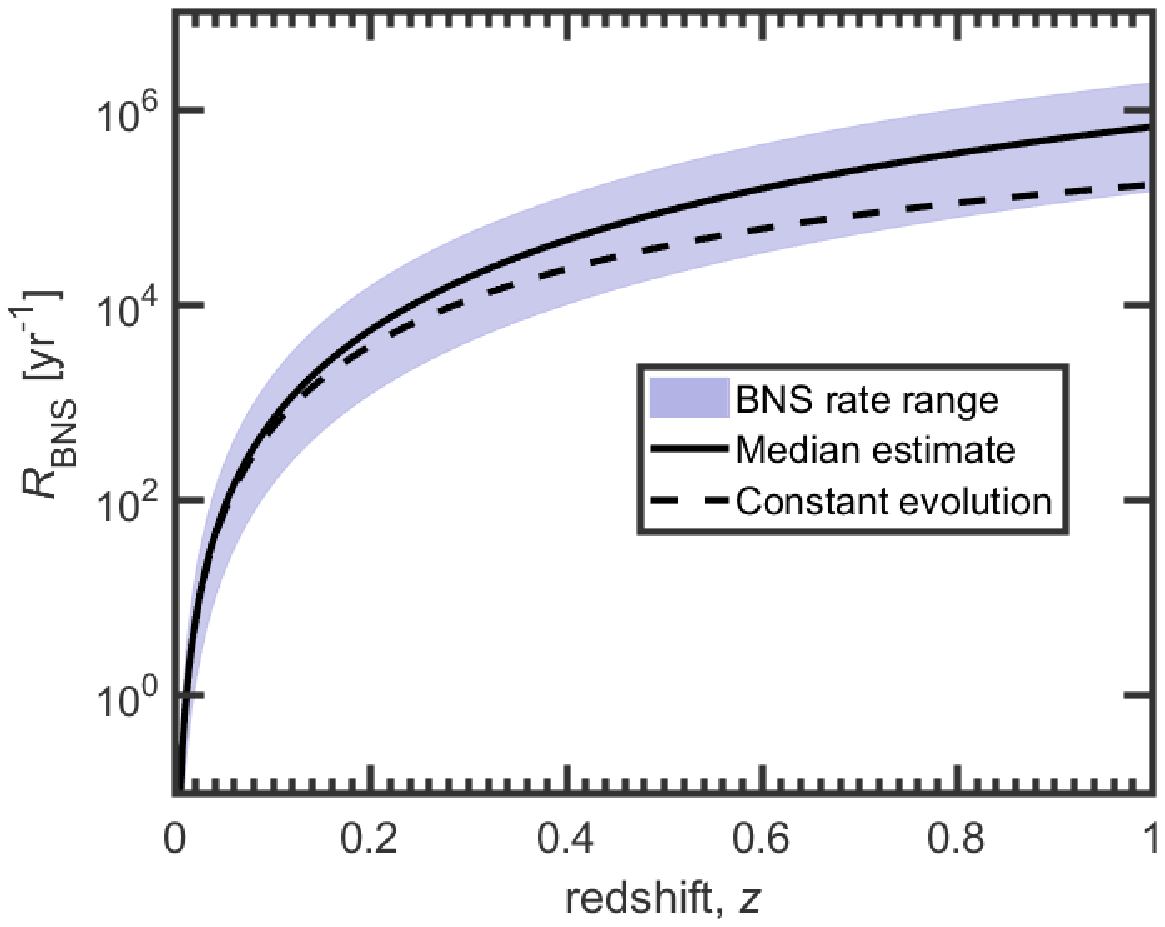}
 \caption{The cumulative all sky rate of BNS mergers based on the the rate estimates of $1540^{+3200}_{-1220}$ \rate yr$^{-1}$ \citep{LSC_BNS_2017PhRvL}. The solid and dashed curves show the integrated rates assuming the model described in section 2.1 and assuming a constant source rate evolution respectively. We find a difference of around a factor 2 at around $z \sim 0.4$ between these two models. The 90\% credible intervals on the intrinsic BNS rate are shown by the shaded band.
 }
  \label{fig:dr}
\end{figure}

Following the procedure of \citet{Regimbau2009PhRvD}, we model the BNS merger evolution by combining the formation rate $R_{\rm F}(z)$ with the delay time distribution $P(t_{\rm d})$:
\begin{equation}\label{eq_bns_rate}
    R_{\mathrm{BNS}}(z)= \int_{t_{\mathrm{min}}}^{t_{\mathrm{max}}} R_{\mathrm{F}}(z_{\mathrm{f}}) P(t_{\mathrm{d}}) \mathrm{d}t_{\mathrm{d}},
\end{equation}

\noindent where $P(t_{\rm d}) \propto 1/t_{d}$, for $t_{\rm d} > t_{\rm min}$, $t_{\rm min} = 20$\,Myr is the minimum delay time for a BNS system to evolve to merger and $t_{\rm max}$ is the Hubble time. We normalise the merger rate $ R_{\mathrm{BNS}}$ so that it corresponds to the local event rate density (volumetric per unit time) of BNS mergers at $z=0$ \citep{Howell2014MNRAS}. Throughout this study we use the BNS rate density estimates of $1540^{+3200}_{-1220}$ \rate yr$^{-1}$ provided by \citet{LSC_BNS_2017PhRvL}. The differential merger rate, $\mathrm{d}R_{\mathrm{BNS}} /\mathrm{d}z$, which describes the event rate within the redshift shell $z$ to $z+{\mathrm d}z$ is then:
\begin{equation}\label{eq_drdz}
\mathrm{d}R_{\mathrm{BNS}} = \frac{\mathrm{d}V}{\mathrm{d}z} \frac{R_{\rm BNS}(z)}{1+z} \mathrm{d}z \,,
\end{equation}
\noindent where the $(1 + z)$ factor accounts for time dilation of the observed rate through cosmic expansion, converting a source-count to an event rate and the co-moving volume element,
\begin{equation}\label{dvdz}
\frac{\mathrm{d}V}{\mathrm{d}z}= \frac{4\pi c}{H_{0}}\frac{d_\mathrm{\hspace{0.25mm}L}^{\hspace{1.5mm}2}(z)}{(1 +
z)^{\hspace{0.25mm}2}\hspace{0.5mm}h(z)}\,,
\end{equation}

\noindent describes how the number densities of non-evolving objects locked into Hubble flow are constant with redshift. The quantity $d_{\mathrm{L}}(z)$ is the luminosity distance.

The cumulative event rate of BNS mergers throughout the Universe is estimated by integrating equation (\ref{eq_drdz}):
\begin{equation}
R_{\mathrm{BNS}}(<z) = \int_{0}^{z}(\mathrm{d}R_{\mathrm{BNS}} /\mathrm{d}z)\mathrm{d}z\;.\label{eq_Rz}
\end{equation}

While sGRBs are routinely detected at cosmological distances the detection range of present second generation (2G) GW detectors to BNSs is still in the Euclidean regime ($z \lesssim 0.1$). As 2.5G instruments such as Voyager come online the reach will be such that projected estimations of the number counts of detected sources will require a more rigorous treatment of cosmic source rate evolution.

Figure \ref{fig:dr} illustrates this point by plotting the function $R_{\mathrm{BNS}}(z)$ based on the rates given in \citet{LSC_BNS_2017PhRvL} based on the observation of GW170817. The solid curve is based on the median BNS rate value with the 90\% credible region indicated by the shaded regions. The dashed curve shows the cumulative rate assuming a constant evolution with redshift ($R_{\mathrm{BNS}}(z) = 1$); we find that this simplification results in a factor of around 2 difference at $z=0.4$ and a factor of 7 at $z=2$, the relative horizon distances of a 2.5G Voyager-type instrument and a 3G instrument such as ET-D \citep{Punturo_ET_2010}. Although our calculations assume only GW detectors up to 2.5G, we note that to model the observed sGRB fraction in section \ref{sec:Fermi_det_rate} requires a complete source rate evolution beyond the Euclidean regime; we therefore choose to adopt a more complete cosmic source rate evolution framework throughout this study.

To convert from an intrinsic BNS rate to a detected rate one must fold in the sensitivity of the GW instruments. The next section will calculate the form of the GW detector efficiency functions for a range of present GW detectors and future upgrades.

\begin{table}
\begin{center}
\centering
\caption{The distances (redshifts, $z$) corresponding to 50\% and 90\% detection efficiencies by present and upgraded GW observatories. The horizon distances at which only optimally orientated and located sources will be detected is shown as a benchmark. }
\label{tab:table_horizon}
\begin{tabular}{l|ccc}
   Epoch & 90\% distance & 50\% distance & Horizon distance \\
         & Mpc ($z$) & Mpc ($z$) & Mpc ($z$) \\
   \hline
03 (2018-19)  & 32               & 86  ($0.02$) & 270 ($0.06$)\\
  aLIGO design& 48 ($0.01$)  & 129 ($0.03$) & 413 ($0.09$)\\
    A +       & 92 ($0.02$)  & 252 ($0.06$) & 842 ($0.17$)\\
    Voyager   & 225 ($0.05$) & 641 ($0.15$) & 2369 ($0.43$)\\
             \hline
\end{tabular}
\end{center}
\end{table}

\begin{figure}
 \centering
 \includegraphics[scale = 0.65,origin=rl]{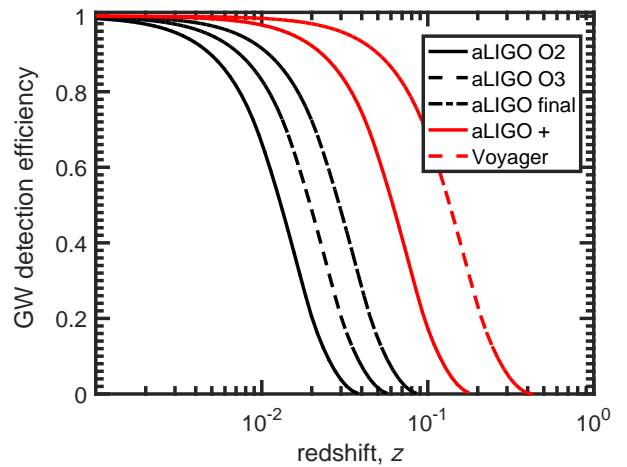}
 \caption{The detection efficiency functions $\Sigma_{\mathrm{GW}}(z)$ for BNS mergers for current and future GW observatories. The curves are produced using the sensitivity noise curves outlined in Appendix A and calculated using the approach of \citet{Belczynski2014ApJ} and \citet{Dominik2015ApJ} who utilised the \emph{projection parameter} described in \citep{FinnChernoff1993PhRvD}.
 }
  \label{fig:efficiency}
\end{figure}

\subsection{BNS detection efficiency}
\label{section:eff_function}

To extrapolate detection rates from the source rate evolution models we require an estimate of the fraction of sources that exceed a GW detection threshold as a function of redshift \citep{Chen2017arXiv}.

In a Euclidean regime one can estimate the number of detections by considering the detection range of an average orientated and located source\footnote{Face on and perpendicular to the detector}. The sensitive range for BNS sources or \textit{SensMon range} is often cited as a measure of the sensitivity of a GW detector and can be estimated by scaling the maximum possible detection distance (for a chosen SNR) or horizon distance by a factor (1/2.26) \citep{Howell2015PASA}. Thus, the sensitive and horizon volumes follow the scaling $V_{\mathrm{sensitive}}/V_{\mathrm{horizon}} \approx (1/2.26)^{3}$. As the reach of future GW detectors (2.5G and beyond) extends beyond the Euclidean regime, where cosmological effects and source rate evolution must be considered, this simple scaling breaks down.

To support sources at cosmological distances, we estimate the fraction of GW sources that exceed a threshold SNR $\rho_{\mathrm{th}}$ as a function of redshift, the GW efficiency function $\Sigma_{\mathrm{GW}}(z)$, following the approach of \citet{Belczynski2014ApJ} and \citet{Dominik2015ApJ} who utilize the projection parameter $\omega$ \citep{FinnChernoff1993PhRvD}. This quantity describes the detector responses for different values of sky location, inclination and polarisation of a GW source. For an optimally-oriented face-on source directly above a GW detector with optimal SNR, $\rho_{\mathrm{opt}}$, $\omega=1$. Conversely, for a poorly located and orientated event $\omega=0$. We can then define $\rho_{\mathrm{th}}$ for any source as it relates to the detector response parameters, or $\rho_{\mathrm{th}} = \omega \rho_{\mathrm{opt}}$. A cumulative distribution function of this quantity $c(\omega)$, which contains all the information of single detector response is provided analytically in \citet{Finn1996PhRvD}.

We construct the function $\Sigma_{\mathrm{GW}}(z)$ by mapping $\omega$ to the distribution $c(\omega)$ through:
\begin{equation}\label{eq_cw}
  \Sigma_{\mathrm{GW}}(z) = c(\omega)=c(\rho_{\mathrm{th}}/\rho_{\mathrm{opt}}(z))\,,
\end{equation}
\noindent \citep{Belczynski2014ApJ,Dominik2015ApJ} for a range of $z$. In equation (\ref{eq_cw}) we note that the value of $z$ at which $\omega=1$ is the horizon redshift $z_{\mathrm{H}}$, which given corresponds to a horizon distance, $d_{\mathrm{H}}$. These quantities represent the absolute detection limit of a given instrument. We assume the standard definition for optimal SNR given as:
\begin{equation}
\rho_{\mathrm{opt}}(z) = 2\left[
  \int_{0}^\infty \, \mathrm{d}f\frac{ \vert\tilde h(f;m_{1},m_{2},s_{1},s_{2},z)\vert^2 }{S_h(f)}\right]^{1/2}\,,
\label{eq:snr}
\end{equation}

\noindent where $\tilde h(f)$ is the BNS waveform in the frequency domain, modeled using the IMRPhenomB waveforms of \citet{ajith_inspiral-merger-ringdown_2011} assuming non-spinning ($s_{1,2}=0$) component masses of $m_{1,2}=$\mbox{[1 + z]1.4\msol} in the detector frame. The function $S_{n}(f)$ is the power spectral density of the the detectors noise for which we the noise curve models described in Appendix A. We further set  $\rho_{\mathrm{th}}$=8 which following standard convention can act as a proxy for a detector network \citep{Abadie2010CQG,Stevenson_2015ApJ,GW150914_AstImp_2016ApJ}.

Figure \ref{fig:efficiency} shows the efficiency functions $\Sigma_{\mathrm{GW}}(z)$ for a range of aLIGO sensitivities and proposed upgrades. For O3 and aLIGO design we assume noise curve with \emph{SensMon ranges} of 120\,Mpc and 173\,Mpc respectively; the sources of our noise curves are described in Appendix A.

Table \ref{tab:table_horizon} outlines the horizon distances (as a benchmark) and following \citep{howell_host_2018} relative distances 50\% and 90\% of BNS mergers will be detected. It is interesting to note that although a future configuration such as Voyager has an optimal reach of 2.4 Gpc, the efficiency at that range will be negligible. To be 90\% confident of detection a source will have to be located at a range of 225 Mpc.

\begin{figure}
  \includegraphics[scale = 0.65,bbllx = 10pt,bblly =20pt, bburx = 500 pt, bbury = 300 pt,origin=lr]{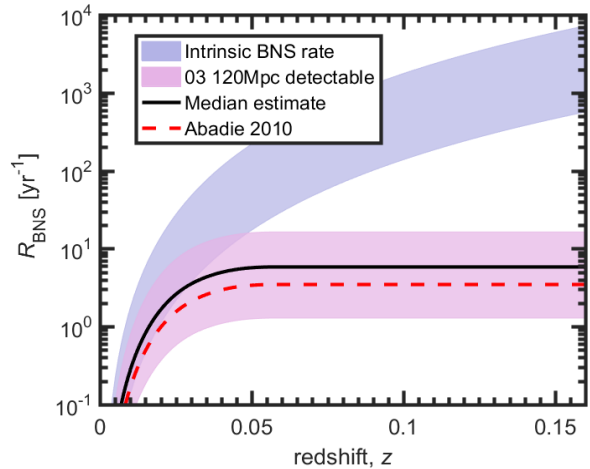}\\
  \caption{The curves above compare the BNS detection rate for the 2019-18 aLIGO/Virgo observation run (O3) with the intrinsic rates based on the estimates of $1540^{+3200}_{-1220}$ \rate yr$^{-1}$ \citep{LSC_BNS_2017PhRvL} determined following the detection of GW170817. The 90\% credible intervals on the intrinsic BNS rate are shown by the purple shaded bands. The equivalent credible intervals on the detection rate are shown by the pink bands; the median estimates are shown by the solid lines.
The dashed line shows the \emph{realistic estimate} of \citet{Abadie2010CQGra} adopted before GW170817.
 }
  \label{fig_BNS_O3_detection_rate}
\end{figure}

\subsection{The BNS event rate}
\label{sec:bns_event_rates}
Given the GW detector efficiency functions $\Sigma_{\mathrm{GW}}(z)$ one can now scale the intrinsic rate by the performance of the GW instrument to yield an expected detection rate for present and future observing epochs. This can be formally expressed through an extension of equation\,(\ref{eq_Rz}):
\begin{equation}
R_{\mathrm{GW}}(z) = R_{\mathrm{BNS}}(z)\, \Sigma_{\mathrm{GW}}(z)\, \Omega_{\mathrm{GW}}\,,
\label{eq_Rz_GW}
\end{equation}

\noindent scaling by $\Sigma_{\mathrm{GW}}(z)$, the GW detector efficiency and $\Omega_{\mathrm{GW}}$, the duty cycle of a typical observing run. We use $\Omega_{\mathrm{GW}} = 0.5$ which is the double coincidence duty cycle of aLIGO over the 2016-17 (O2) observing run\footnote{Taken from the internal document \url{https://dcc.ligo.org/DocDB/0153/G1801329}} and we assume this value for all future observing runs. In the following section we estimate the value of $\Sigma_{\mathrm{GW}}(z)$ for a range of planned aLIGO/Virgo observation runs and upgrades. The noise curves we use for this analysis are documented in Appendix A.

Table \ref{table_BNSRates} shows the predicted rate of detectible BNS systems per year during a range of future observing runs. We check these estimates by comparing the number of detections predicted in O2 by our model to the single event detected, GW170817. Using a representative curve for O2 described in the Appendix A, we find a range of 0.2-2.5 events per year with a median of 0.8 for a 9 month observation in agreement O2. Given our model, the Poisson probability of at least one event detected within 42.5\,Mpc during O2 is around 11\%.

Figure \ref{fig_BNS_O3_detection_rate} shows the projected O3 BNS detection rate as a function of redshift in comparison with the intrinsic population. We see that the detected rate becomes asymptotic at around $z\sim0.05$ to a value $5.3 ^{+11.1}_{-4.2}$yr$^{-1}$ (1.1-16.4 yr$^{-1}$). Table \ref{table_BNSRates} provides estimates for aLIGO at design sensitivity and shows that we can expect between 3.7--55 events a year during this period with a median expected value of around 18 yr$^{-1}$.

\begin{figure}
 \includegraphics[scale = 0.6,bbllx = 0pt,bblly =0pt, bburx = 500 pt, bbury = 300 pt,origin=lr]{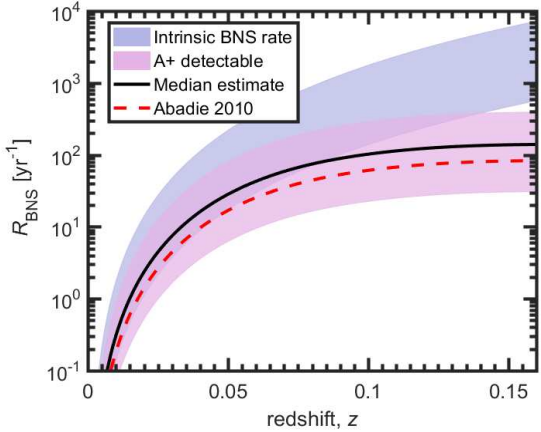}\\
  \caption{As for Fig \ref{fig_BNS_O3_detection_rate} but with the A+ configuration. For A+ the detection efficiency is 50\% at around $z\sim0.06$. The detection curve becomes asymptotic at around $z=0.14$ corresponding to around a 4\% detection efficiency at this range.}
  \label{fig_BNS_Aplus_detection_rate}
\end{figure}

Looking further ahead, Figure \ref{fig_BNS_Aplus_detection_rate} shows the BNS detection rate expected during the A+ upgrade. The BNS rate becomes asymptotic at $z\sim0.15$ with an expected event rate of $128.2 ^{+266.3}_{-101.5}$; the range of this upgraded instrument will permit valuable in depth studies of the BNS population. When the proposed Voyager instrument comes on line, as indicated by Table \ref{table_BNSRates}, estimates of $1822.5 ^{+3787.1}_{-1443.8}$ will allow one to perform significantly detailed studies of distribution of component masses in binary systems and evolutionary effects \citep{finn_binary_1996}.

\begin{table}
\begin{center}
\centering
\caption{LIGO BNS rates calculated assuming a 50\% double coincidence duty cycle. \dag - 9 month estimate \ddag best estimated observing schedule.}
\label{table_BNSRates}
\begin{tabular}{cccccc}
\hline
\R     O2 \dag    & O3         & aLIGO  & A+ \ddag             & Voyager \ddag \\
\R      (2016-17) & (2019-20)  & 2021-  & 2024-  & 2030-  \\ \hline
\R   $0.8 ^{+1.7}_{-0.6}$ & $5.3 ^{+11.1}_{-4.2}$ & $17.9 ^{+37.2}_{-14.2}$ & $128.2 ^{+266.3}_{-101.5}$ & $1822.5 ^{+3787.1}_{-1443.8}$\\
\\ \hline
\end{tabular}
\end{center}
\end{table}

\subsection{The sGRB detection rate of Fermi}
\label{sec:Fermi_det_rate}
Given the intrinsic BNS rate evolution model $R_{\mathrm{BNS}}(z)$, and assuming that all BNS mergers can produce a sGRB one can calculate a detection rate for Fermi-GBM by scaling $R_{\mathrm{BNS}}(z)$ by the sGRB detection efficiency model determined in section \ref{sec:Fermi_efficiency_function}:
\begin{equation}
R_{\mathrm{GRB}}(z) = R_{\mathrm{BNS}}(z)\, \Sigma_{\mathrm{GRB}}(z)\, \Lambda_{\mathrm{GRB}} f_{\mathrm{b}} \;.\label{eq_Rz_GRB}
\end{equation}
\noindent Here, we also introduce the total time-averaged observable sky fraction of the Fermi-GBM which is given as $\Lambda_{\mathrm{GRB}}$=0.60 \citep{Burns2016ApJ}.

\begin{figure}
 \includegraphics[scale = 0.64,origin=rl]{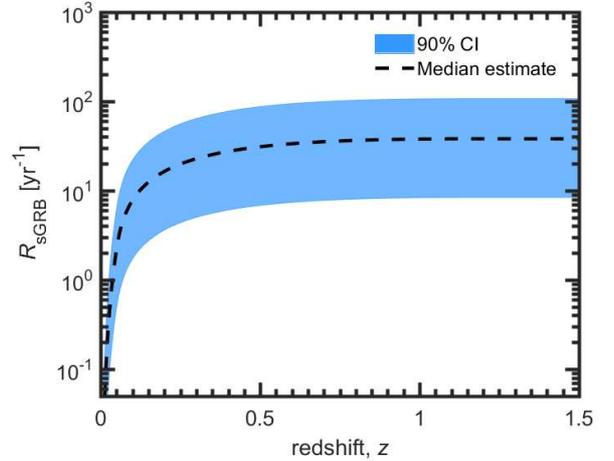}\\
  \caption{The detection rate of sGRBs by Fermi-GBM based on the efficiency function shown in Fig \ref{fig:structured_profile} calculated using the inferred parameters for the structured jet profile of GRB170817A. The detected proportion of sGRBs flattens at around $z\sim 0.6$ due to flux limit of the instrument. The median detection rate of around 39\,yr$^{-1}$ is consistent with the average number recorded by Fermi-GBM since 2008. }
  \label{fig_fermi_detection_rate}
\end{figure}

Figure \ref{fig_fermi_detection_rate} plots the function $R_{\mathrm{GRB}}(z)$ and shows that the detection fraction becomes asymptotical at around $z=0.8$-1 as detections of sGRBs become rarer. We obtain a Fermi-GBM detection rate of $38.74 ^{+80.50}_{-30.69}$, compatible with the expected rate of 39.8 sGRBs/yr detected by Fermi-GBM since 2008\footnote{Calculated from 10.2 years of data using the Fermi-GBM burst catalogue at \url{https://heasarc.gsfc.nasa.gov/W3Browse/fermi/fermigbrst.html}}. In \citet{LSC_GW_GRB_2017ApJ} the redshift distribution predicted by different forms of the LF was scaled to produce the observed Fermi-GBM detection rate; however, in this paper we find that the Fermi-GBM detection rate is a useful cross-test and suggest it could be used as a prior in further studies.

One can constrain the upper limit on the first integral of equation\,(\ref{eq_grb_eff}) to determine the fraction of detections observed within an arbitrary opening angle. If we assume a mean half opening angle of $<$10\deg$ >$ ($f_{\mathrm{b}} = 0.015$) \citep{Berger2014ARA&A} we find a detection rate of around $27.83 ^{+57.84}_{-22.05}$ which is over 70\% of the total detection fraction. Inspection of Fig \ref{fig:fermi_max_angle_z} suggests such a proportion of Fermi-GBM sGRBs are detected from sources outside $z\sim0.4$.

\subsection{Assessing the possibility of wider-angled emissions}
\label{sec_using_wider angled_model}
\begin{figure}
 \centering
 \includegraphics[scale = 0.66,origin=rl]{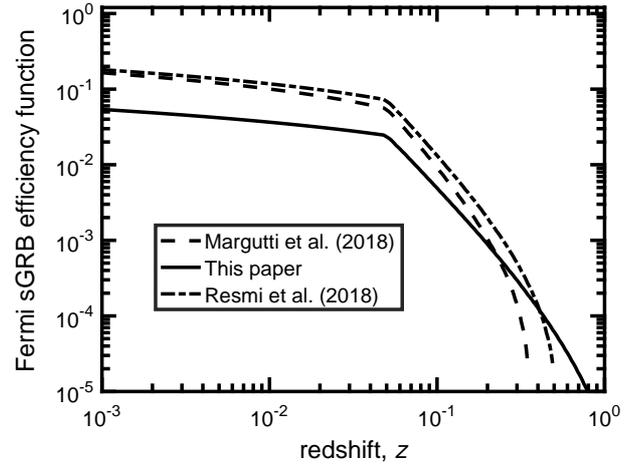}
 \caption{The Fermi-GBM detection efficiency functions for structured jets with a jet profiles inferred for GRB170817A. The efficiency model adopted in this paper is shown by the solid line and compared with two models predicting wider-angled emissions: the model-M of \citet[][]{margutti_binary_2018} is shown by the dashed curve and the model-R of \citet[][]{resmi_low_2018} is indicated by the dot-dashed curve.}
  \label{fig:sgrb_efficiency_2}
\end{figure}

\begin{figure}
 \centering
 \includegraphics[scale = 0.66,origin=rl]{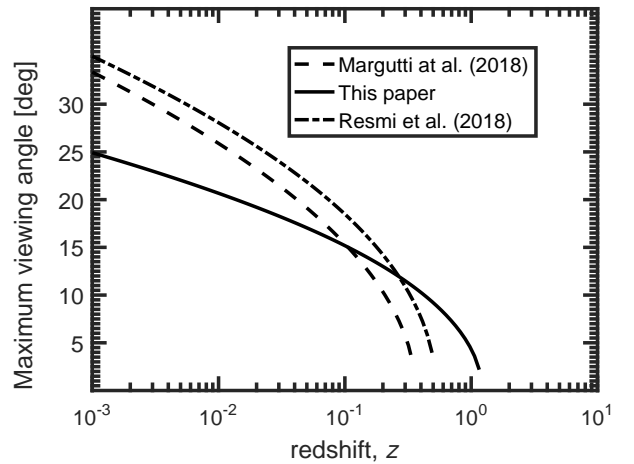}
 \caption{As for Fig.\ref{fig:sgrb_efficiency_2} but showing the relative maximum observable viewing angle as a function of redshift assuming the different structured jet profiles.}
  \label{fig:fermi_max_angle_z_2}
\end{figure}

Although the \citet{Mooley_2018} observations constrain an observing angle $\sim 20$\deg (the lower end of Figure \ref{fig:structured_models}) for GRB170817A, it is worth looking at the efficiency function produced by a models suggesting wider viewing angles closer to 30\deg. In this section we use the models of \citet[][model-M]{margutti_binary_2018} and \citet[][model-R]{resmi_low_2018} and the relative parameters for $\theta_{\mathrm{c}}$ and $\theta_{\mathrm{v}}$ provided in Table \ref{table_sj_observations}. We iterate through the range of values of $L_{\mathrm{c}}$ that agree with the observed luminosity $L_{\mathrm{obs}}$ and $\theta_{\mathrm{v}}$. We find a value of $L_{\mathrm{c}}\sim 7 \times 10^{50}/\Omega$ erg s$^{-1}$ sr$^{-1}$ for model-M and $L_{\mathrm{c}}\sim 1.5 \times 10^{51}/\Omega$ erg s$^{-1}$ sr$^{-1}$ for model-R. Repeating the procedure of section \ref{sec:Fermi_efficiency_function} the resulting efficiency functions are shown in Figure \ref{fig:sgrb_efficiency_2} and the maximum observable viewing angle as a function of redshift in Figure \ref{fig:fermi_max_angle_z_2}.

Figure \ref{fig:sgrb_efficiency_2} shows there is would be a greater contribution of sGRBs at lower redshift for both model-M and model-R in comparison to our model. Model-R predicts a quite similar contribution within $z\sim0.05$ but predicts a slightly greater fraction at higher-$z$ although less than the model adopted in this paper. Figure \ref{fig:fermi_max_angle_z_2} shows that both Model-M and Model-R suggest a low-$z$ contribution dominated by wide-angled emissionsup to opening angles of 35\deg.

Figure \ref{fig:fermi_max_angle_z_2} shows the maximum viewing angle for the range of models. The plot seems to suggest a maximum distance corresponding to $z=0.2$ and $z=0.4$ for sGRBs viewed near the core (assuming an arbitrary value of half opening angle 10\deg) for model-M and model-R respectively.

\begin{figure}
 \centering
 \includegraphics[scale = 0.66,origin=rl]{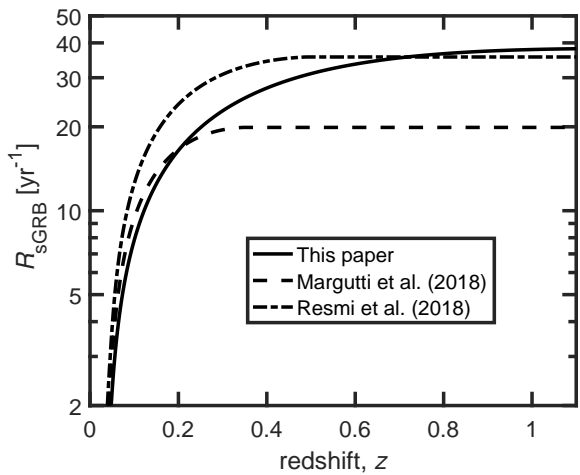}
 \caption{As for Fig.\ref{fig:sgrb_efficiency_2} but showing the Fermi-GBM detection rates as a function of redshift predicted by the efficiency functions of Fig.\ref{fig:sgrb_efficiency_2}.}
  \label{fig:fermi_rate_2}
\end{figure}

Figure \ref{fig:fermi_rate_2} plots the relative detection rates as a function of redshift. Repeating the analysis of section \ref{sec:Fermi_det_rate} we find a rate of $17.69 ^{+36.77}_{-14.02}$ observed by Fermi-GBM for model-M of which 55\% are detected from emissions within a 10\deg opening angle.
For Model-R the predictions for Fermi-GBM are $33.00 ^{+68.57}_{-26.14}$ yr$^{-1}$; nearer to the observed number, with 48\% detected from emissions within a 10\deg opening angle. We see that model-M and model-R produce very few detections outside $z=0.3$ and $z=0.5$ respectively, as indicated by the redshift at which the curves become asymptotic.

Both these model predict a greater fraction of detections at low-$z$ from wider-angled emissions. The model-R seems to predict a Fermi-GBM detection rate closer to the observed value of around 40\,yr$^{-1}$ and would suggest an optimistic number of joint GW/sGRB detections. However, it is at odds with the median redshift of sGRBs $z \sim0.5$ \citep{Berger2014ARA&A,fong_decade_2015}. For example, there have been Fermi-GBM triggers coincident with \emph{Swift} allowing redshift determinations via follow-up campaigns close to $z=1$ at which the detection rate curve of Figure \ref{fig_fermi_detection_rate} flattens out: two coincident bursts on the \emph{Swift} online database are: GRB090510 \citep[at $z$=0.903;][]{Ackermann2010ApJa}
and GRB131004A (at $z$=0.71; Chornock et al. GCN 15307). The former burst was, however, a particularly bright and hard event with an additional hard power-law component up to the GeV range.

Given the sGRB efficiency function used in this paper is based on the structured jet profile of a single event, GRB170817A, it is reasonable to suggest that a range of profiles will be determined for joint GW/GRB detections in forthcoming years. However, with a sample of one event, we argue that the derived efficiency function of section \ref{sec:Fermi_efficiency_function} is consistent with the Fermi-GBM sample of sGRBs and can provide a realistic constraint of joint GW/GRB rates.

\section{Joint GW/sGRB detection rates}
To calculate a joint GW/sGRB detection rate we combine the scaling factors of equations\,(\ref{eq_Rz_GW}) and (\ref{eq_Rz_GRB}) to yield:
\begin{equation}
R_{\mathrm{GW/GRB}}(z) = R_{\mathrm{BNS}}(z)\, \Sigma_{\mathrm{GW}}(z)\, \Sigma_{\mathrm{F}}(z)\, \Omega_{\mathrm{GW}}  \Lambda_{\mathrm{F}} f_{\mathrm{b}}\,.\label{eq_Rz_GWGRB}
\end{equation}

For a joint GW/sGRB detection the more constrained temporal and spatial constraints essentially lower the SNR threshold for the GW detection search; thus the detection range is increased. Therefore, the joint detection efficiencies will differ with an EM counterpart; i.e. threshold SNR $\rho_{\mathrm{t}}$ of equation \ref{eq_cw} should be recalculated to reflect the situation where there is an EM counterpart. To do this we use the relation between the SNR for triggered, $\rho_{\mathrm{T}}$ and un-triggered searches $\rho_{\mathrm{U}}$ given as \citep{Bartos2015PhRvL, patricelli_prospects_2016}:

\begin{equation}\label{eq_snr_joint_GWGRB}
  \rho_{\mathrm{T}} = \sqrt{2 \mathrm{log} \left[ \mathrm{exp}\left(\frac{\rho_{\mathrm{U}}^{2}}{2} \right) \frac{\Omega_{\mathrm{T}}^{2}\times t_{\mathrm{obs},\mathrm{T}}}{\Omega_{\mathrm{U}}^{2}\times t_{\mathrm{obs},\mathrm{U}}} \right]    }
\end{equation}

\noindent where $\Omega_{\mathrm{T,U}}$ and $t_{\mathrm{obs},\mathrm{T,U}}$ are the relative sky regions and observation durations respectively.

Following \citet{patricelli_prospects_2016} we take $\Omega_{\mathrm{T}}$=100 deg$^{2}$,  $\Omega_{\mathrm{U}}$=40000 deg$^{2}$ and $t_{\mathrm{obs},\mathrm{U}}$ = 1 yr. For $t_{\mathrm{obs},\mathrm{T}}$ we assume $\Delta t \times N_{\mathrm{sGRB}}$ where $\Delta t$ is the on-source window which we take as 6s and $N_{\mathrm{sGRB}}$ is the number of expected sGRB detections per year within the GW detector horizon.

\begin{figure}
 \centering
 \includegraphics[scale = 0.71,origin=rl]{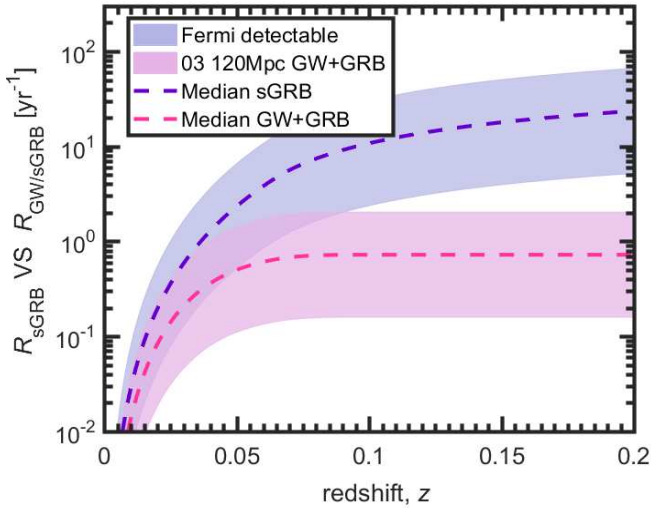}
 \caption{The joint sGRB-BNS detection rates for the O3 observation run (2019-20) is compared to the sGRB detection rate of Fermi-GBM. This plot suggests a reasonable chance of another joint GW/sGRB detection during this run.
 }
  \label{fig:sgrb_grb_O3}
\end{figure}

Figure \ref{fig:sgrb_grb_O3} shows the joint GW/sGRB detection rate as a function of redshift for O3 in comparison with the corresponding detection rate of Fermi-GBM. We see that the coincident rate becomes asymptotic at around $z\sim0.07$ to a value $0.58 ^{+1.21}_{-0.46}$; comparison with Fig.\ref{fig:fermi_max_angle_z} suggests that at this distance the sGRB counterparts would most likely be from wider-angled emissions.

\begin{figure}
 \centering
 \includegraphics[scale = 0.71,origin=rl]{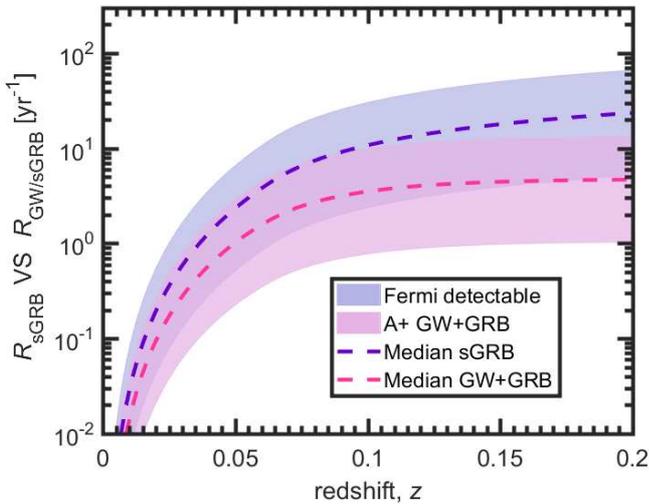}
 \caption{As for Figure \ref{fig:sgrb_grb_O3} but based on the A+ aLIGO upgrade. We see that 1-3 joint detections per year ar possible out to $z \sim 0.15$. This plot shows that a greater fraction of the Fermi-GBM sGRBs will have GW counterparts.
 }
  \label{fig:sgrb_grb_APlus}
\end{figure}

Figure \ref{fig:sgrb_grb_APlus} shows the joint GW/sGRB detection rate expected during the A+ upgrade. The joint-rate becomes asymptotic at $z\sim0.13$ at a value of $3.10 ^{+6.45}_{-2.46}$yr$^{-1}$ during this observation run; the increased range of A+ suggests a greater probability of an emission closer to the core (as evidenced by Fig.\ref{fig:fermi_max_angle_z} ); the implications will be discussed later in this section.

Table \ref{tab_joint_gw_grb} shows full range of joint sGRB/GW detection rates expected for present and future GW observation runs. We note that our calculation for the 9 months of O2 of $0.14 ^{+0.30}_{-0.11}$\rate suggests around a 1 in 5 chance of recording a coincident GW/sGRB event based on the GRB170817A jet structure profile; however, given our modeling recovers the O2 BNS rate (section \ref{sec:bns_event_rates}) and is in agreement with the Fermi-GBM detection rate (section \ref{sec:Fermi_det_rate}), we suggest that this estimate is in agreement with the probability of observing GW170817/GRB170817A. We see that projecting further ahead, a future Voyager detector with improved sensitivity could provide 10-20 joint GW/sGRB events each year; such numbers would start to enable in depth studies of the sGRB population.

The range of our joint GW/sGRB rates for the 2019-20 observing run (O3) are 0.1 - 1.8 detections per year and around 0.3--4 per year at design sensitivity. These estimates more optimistic in the higher range than those given in \citep{LSC_GW_GRB_2017ApJ} of 0.1--1.4 yr$^{-1}$ (O3) and 0.3--1.7 yr$^{-1}$ (design); in this study the luminosity function was extended in the low end to $L_{\mathrm{min}} = 10^{47}$\,\lum so did not consider wider-angled emissions. \citet{clark_prospects_2015, LSC_GW_GRB_2017ApJ} produces estimates that convert to around  0.006--0.2 and 0.1--0.7 for 1 year runs at aLIGO sensitivities of 120 Mpc and 170 Mpc respectively; equivalent to O3 and design sensitivity. Our estimates are far more optimistic in this case as this study only assumed a top-hat model.

The second row of Table \ref{tab_joint_gw_grb} shows the percentage of BNS events that would be accompanied by a sGRB during each upgrade of aLIGO. It is interesting to note that the percentage decreases with each subsequent upgrade. This trend is surprising until one considers that the more probable detections originate from wider angled emissions which are less accessible at greater distances due to the flux limited sensitivity of a GRB detector. In the future, joint sGRB/GW detections may be limited by the GRB detection range to wider-angled emissions rather than the GW detection range. This fact was evidenced in O2 by the maximum detection range of GRB170817A being less than the sensitive distance of the aLIGO instruments (see also section \ref{sec:jet_profile_inference}).

Table\,\ref{tab_joint_gw_grb_10deg} shows the joint sGRB/GW detection rates of events with emissions from within an opening angle of 10\deg; as in section \ref{sec:Fermi_det_rate} one can assume that imposing this limit will correspond to the brightest part of a structured jet close to the core. This is important as one would expect these emissions close to the jet-axis to be accompanied by an early-time X-ray afterglow that a detector such as \emph{Swift} with rapid slew capabilities could catch on source. We note that \emph{Swift} could be on source within order hours post trigger for a burst accompanied by a LVC trigger and within around 25 minutes if specific coordinates of an interesting source are provided\citep{Tohuvavohu2017,Tohuvavohuinprep}.

Observations of plateau features in early X-ray afterglows can provide valuable information on the post-merger remnant \citep{Rowlinson2010MNRAS,Rowlinson2013MNRAS} which can be combined with GW observations to place constraints on the nuclear equation of state \citep{Lasky2014PhRvD,2016MNRAS.458.1660L} or ellipticity \citep{2018PhRvD..98d3011S}.

\begin{table}
\begin{center}
\centering
\caption{Joint sGRB/GW rates calculated assuming a double coincidence 50\% duty cycle for two aLIGO type detectors and a Fermi-GBM instrument with a 60\% duty cycle. The bottom row shows the percentage of BNS detections that are detected with a sGRB; we see that the fraction decreases with GW detector sensitivity. \dag = 9 month estimate; \ddag = best estimated observing schedule.}
\label{tab_joint_gw_grb}
\begin{tabular}{ccccc}
\hline
\R     O2 \dag    & O3         & aLIGO  & A+ \ddag             & Voyager \ddag \\
\R      (2016-17) & (2019-20)  & 2021-  & 2024-  & 2030-  \\ \hline
\R
$0.14 ^{+0.30}_{-0.11}$& $0.58 ^{+1.21}_{-0.46}$ & $1.23 ^{+2.55}_{-0.97}$& $3.10 ^{+6.45}_{-2.46}$ & $7.53 ^{+15.66}_{-5.97}$ \\ 
- & 11\% & 7\%  & 2\%  & 0.4\%
\\ \hline
\end{tabular}
\end{center}
\end{table}

\begin{table}
\begin{center}
\centering
\caption{As for Table \ref{tab_joint_gw_grb} but for events detected within an arbitrary opening angle of 10\deg of the jet axis. Such events could provide valuable early-time X-ray afterglows assuming a \emph{Swift} type detector can get on source in sufficient time.}
\label{tab_joint_gw_grb_10deg}
\begin{tabular}{ccccc}
\hline
\R     O2 \dag    & O3         & aLIGO  & A+ \ddag             & Voyager \ddag \\
\R      (2016-17) & (2019-20)  & 2021-  & 2024-  & 2030-  \\ \hline
\R
$0.05 ^{+0.11}_{-0.04}$& $0.21 ^{+0.44}_{-0.17}$& $0.46 ^{+0.96}_{-0.37}$& $1.29 ^{+2.68}_{-1.02}$ & $3.85 ^{+8.00}_{-3.05}$\\ 
- & 4\% & 3\%  & 1\%  & 0.2\%
\\ \hline
\end{tabular}
\end{center}
\end{table}

\section{Conclusions}

In this paper we have described a framework to determine joint sGRB/GW detection rates assuming a structured jet profile for GRB170817A. We have assumed that all BNSs can produce a sGRB and have based our modeling around the BNS rates determined from the O2 observation run. To convert from intrinsic BNS rates to detection rates we have folded in both GW and sGRB detection efficiency functions. We have assumed Fermi-GBM for our GRB instrument and calculated an efficiency function assuming a well cited parameters for the sGRB LF and its parameters \citep{WandermanPiran2015MNRAS}\footnote{We have independently verified the form of the LF used here through $\mathrm{log}\,N-\mathrm{log}\,P$ fitting to Fermi-GBM data.}, folding in the geometric dependence that results from a structured jet.

To verify of our modelling, we have shown that our derived BNS detection rate is consistent with a single detection in O2 and furthermore, that the sGRB detection rate is consistent with Fermi-GBM.

We have determined a structured jet profile using Bayesian inference based on the data from early and late EM observations. An important prior used in this analysis are the VLBI observations of the super-luminal motion of GRB170817A provided by \citet{Mooley_2018} that indicated a successful jet with a viewing angle of 20$\pm$5\deg \citep[see also][]{Alexander_2018ApJ}.

We have investigated the use of wider-angled emission models and present our case in Section \ref{sec_using_wider angled_model}. Although wider angled models predict a greater contribution of sGRBs at lower redshift (see Figure \ref{fig:sgrb_efficiency_2}) the event rate density is lower at smaller cosmological volumes resulting in a Fermi-GBM detection rate much below what is observed. Additionally, the wider angled models predict few events higher than $z \sim 0.4$; this is in tension with Fermi-GBM sGRBs observed in coincidence with \emph{Swift} and thus having redshift estimates (see section \ref{sec:Fermi_det_rate}).

We note that a recent study conducted by \citet{Bartos2018} also considered the implications of joint sGRB/GW detections from structured jets using a numerical approach based on the model of \citet{margutti_binary_2018} that predicts a viewing angle for GRB170817A of 27\deg (see also section \ref{sec_using_wider angled_model}). This study finds a larger fraction of joint sGRB/GW detections due to two main factors: the structured jet model used and a relatively low fluence threshold\footnote{The fluence threshold is $2.5 \times 10^{-8}$; we find that only only 0.5\% of sGRBs are lower than this value in the 50-300\,keV band}. It is clear that present studies are model dependent, and we suggest that \citet{Bartos2018} represents a more optimistic joint detection scenario. We look forward to future joint sGRB/GW detections converging towards a universal structured jet-profile or alternatively, a more complex suite of physical models.

We find that the fraction of coincident GW/sGRB events will decrease as the sensitivity of GW detectors increase; our results in Table \ref{tab_joint_gw_grb} show that the joint detection percentage of BNS will be 11\% during O3, decreasing to less than 1\% by the time of the Voyager aLIGO upgrade. This projection is due to only the rarer brighter emissions from closer to the jet axis of a structured jet being detectable by Fermi-GBM as the GW detection range increases (see Fig. \ref{fig:fermi_max_angle_z})

Our modeling allows us to constrain the number of events that may be observed within an arbitrary viewing angle. This has enabled us to produce joint GW/sGRB detection rates and predict for coincident events observed observed close to the jet core; such events would be valuable, allowing an early X-ray emission could be observed by a fast slewing detector such as \emph{Swift}. For example, an X-ray plateau \citep{Rowlinson2010MNRAS,Rowlinson2013MNRAS} coupled with GW data could provide valuable information on the post-merger remnant \citep{Lasky2014PhRvD,2018PhRvD..98d3011S}. Our modelling suggests that 70\% of Fermi-GBM detections are from sGRBs viewed within a half opening angle of 10\deg, most likely from sources $z\gtrsim0.4$. Thus, such observations will require more sensitive GW detectors; we find that around 1-4 such joint observations could be achieved during A+ and up to 10 during Voyager.

A factor not considered in this study that could increase the number of joint sGRB/GW detections are ground based searches for subthreshold events. Such events, which are intrinsically weak, distant or have unfavourable viewing geometry, fail to achieve the triggering requirement of the Fermi-GBM\footnote{Triggering at least two of the 14 Fermi-GBM detectors.} \citep{FermiSubThresh_2018}. These offline methods can provide an additional 80 sGRB candidates per year and have been verified by finding 9 of 11 sGRBs detected by \emph{Swift} but not initially detected by the GBM, despite being within the observational window of the instrument. Although the percentage of subthreshold events that would occur within a GW detection volume is uncertain, the detection rates provided in this paper could increase though the detection of intrinsically weaker events or from dimmer emissions resulting from events viewed at slightly wider viewing angles than suggested in this study.

Motivated by the firm observation of GW170817/GRB170817A, this study has only assumed BNSs as progenitor systems to sGRBs. Mergers of
BH/NS systems have also been predicted to produce sGRBs if the conditions are favorable for NS disruption outside the BH horizon and thus the formation of an accretion disk. Required conditions include the BHs having low enough mass and prograde spins relative to the system orbit\citep{Bhattacharya_2018,DesaiMetzger2018,Shapiro2017PhRvD}\footnote{Misaligned
black-hole spins result in a larger fraction of that material in the tidal tail rather than
forming an accretion disk.}. Interestingly, sGRBs with extended emissions (sGRB-EE), bursts with an initial short hard spike ( $< 2$\,s) followed by a faint softer emission ($\gtrsim 100$\,s) \citep{norris_06,sakamoto_BAT2_2011,Howell2013MNRAS} have been suggested to result from BH/NS star mergers; the extended emissions could be produced by fallback accretion \citep{DesaiMetzger2018}. Typically, this sub-category of sGRBs are predicted to account for around 10\% of the sGRB population \citep{Lien2016ApJ}. Whether BH/NS mergers, which would have greater GW detection ranges than BNS mergers, could produce joint sGRB/GW detections will depend on factors such as their intrinsic rate \citep{2018arXiv181112907T}, the jet profiles and energetics; these factors are uncertain. If future GW observations are able to probe the BH/NS population, constraints from the GW and EM domains can accelerate our knowledge on this possible formation channel of sGRBs.


In conclusion, the framework we have presented here highlights that joint GW/sGRB rates are limited by the GRB instrument rather
than the GW detectors. For the remainder of the era of advanced GW interferometers, joint GW/sGRB detections will most likely
be from sGRBs observed at wide-viewing angles. At the increased GW detection ranges of planned or proposed detector upgrades,
flux-limited GRB detectors will be unable to detect the wider angled emissions; sGRB detections will start to be dominated by
emissions closer to the jet axis. This pattern will continue into the era of 3G interferometers such as ET and Cosmic Explorer.
Thus, consideration of the potential GRB/X-ray instruments that will be available in the future \citep[e.g.THESEUS;][]{stratta_theseus_2018} will become increasingly important for studies of
GRBs.

\begin{appendix}

\section{GW Interferometer Sensitivity Noise curves}
\label{sec_noise_curves}
To estimate the performance of O2 (2016-2017) we use the mid (low) projected noise curve of \citet{abbott_prospects_2018} provided in \url{https://dcc.ligo.org/LIGO-T1200307} for which the estimated range is around 80 Mpc; we scale this curve to provide a range of around 70 Mpc which was around the average performance for H1 during O2 during the 9 months of observation. Although L1 archived a fairly consistent range of around 90Mpc with best performance over 100Mpc (see \url{https://ldas-jobs.ligo.caltech.edu/~detchar/summary}\\\url{/1126623617-1136649617/range/} we conservatively assume here the performance of the least sensitive instrument. For our O3 (2019-20) estimates we use the late (low) curve taken from the same resources as above. For aLIGO design we use the updated curve provided at \url{https://dcc.ligo.org/LIGO-T1800044} for which the projected range is 173 Mpc. For the A+ noise curve we use the model found at \url{https://dcc.ligo.org/LIGO-T1800042} and for Voyager the version found at \url{https://dcc.ligo.org/LIGO-T1500293}.
We note that optimal SNR values, horizon distances and detection ranges can be checked using the \verb"GWINC"  Gravitational Wave Interferometer Noise Calculator which can be obtained at \url{https://git.ligo.org/gwinc/matgwinc}.

\section{Top-hat GRB efficiency function}
\label{sec_top_hat_eff}
In the top hat model, the efficiency function takes the form:
\begin{equation}
\Sigma_{\mathrm{TH}}(z) = f_{\mathrm{b}}\int_{L_{\mathrm{min}}(\mathrm{P_{T}},z) }^{L_{\mathrm{max}} }\!\! \acute{\Phi}(L) \mathrm{d}L,
\label{eq_top_hat_grb_eff}
\end{equation}
\noindent where $f_{\mathrm{b}}=\langle [1 - \mathrm{cos}\, \theta_{\mathrm{j}}]^{-1} \rangle $ is the collimation factor, or beaming fraction.

\end{appendix}
\section{ACKNOWLEDGMENTS}
Parts of this research were conducted by the Australian Research Council Centre of Excellence for Gravitational Wave Discovery (OzGrav), through project number CE170100004. EJH acknowledges support from a Australian Research Council DECRA Fellowship (DE170100891). The authors gratefully thank Raymond Frey (University of Oregon), the assigned reviewer for the LIGO Scientific Collaboration, for conducting a thorough review of the manuscript which included a number of insightful suggestions. We are also particularly grateful to Giulia Stratta (INAF/OAS) who provided some useful comments to improve the manuscript. We also thank Aaron Tohuvavohu of the Neil Gehrels Swift Observatory (\emph{Swift}) for useful discussions on the follow-up capabilities of \emph{Swift}. Finally, we thank the anonymous referee for a thorough reading of the manuscript and for providing a number of erudite suggestions that have enriched the paper. This paper has LIGO document ID P1800322.
\bibliographystyle{mnras}

\end{document}